\documentstyle[12pt]{article}
\begin{document}
\baselineskip=15pt
\hoffset=-10 mm
%\voffset=-32 mm

%---------------------------------------------------------------------------
\renewcommand{\baselinestretch}{1.5}
\renewcommand{\theequation}{\arabic{equation}}
%----------------------------------------------------------------------------

\title{{Charge Fluctuation Forces Between Stiff }
\\{Polyelectrolytes in Salt Solution:}
\\{Pairwise Summability Re-examined}}
\author{ R. Podgornik \thanks{On leave from Dept. of Physics,
Faculty of Mathematics and Physics, University of Ljubljana and
Dept. of Theoretical Physics, "J.Stefan" Institute, Ljubljana, 
Slovenia.} and V.A. Parsegian
\\{\sl Laboratory of Structural Biology}
\\{\sl Division of Computer Research and Technology}
\\{\sl National Institutes of Health, Bethesda, MD 20892-5626} }

\date{}
\begin{titlepage}
\maketitle
\begin{abstract}
\baselineskip=8pt
\small We formulate low-frequency charge-fluctuation forces between 
charged cylinders ­ parallel or skewed ­ in salt solution: forces from 
dipolar van der Waals fluctuations and those from the correlated 
monopolar fluctuations of mobile ions.  At high salt concentrations 
forces are exponentially screened.  In low-salt solutions dipolar 
energies go as $R^{-5}$ or $R^{-4}$; monopolar energies vary as 
$R^{-1}$ or $\ln{R}$, where $R$ is the minimal separation between cylinders.  
However, pairwise summability of rod-rod forces is easily violated in 
low-salt conditions.  Perhaps the most important result is not the 
derivation of pair potentials but rather the demonstration that some 
of these expressions may not be used for the very problems that 
originally motivated their derivation.

\centerline{\today }
\end{abstract}
\end{titlepage}

%-------------------------------------------------------------------------
\baselineskip=12pt
\parskip=15pt plus 1pt 
\parindent=40pt
%--------------------------------------------------------------------------

Models of assembly by rod-like particles often require closed-form 
expressions for molecular interaction.  In addition to brute steric 
forces, hydration forces, and electrostatic double layer interactions 
\cite{ref1} between charged rods, there are interactions of correlated 
charge fluctuations \cite{ref2}.  These fluctuations can be either the 
dipolar events of traditional van der Waals forces or monopolar charge 
fluctuations from transient changes in the density of mobile ions 
around a charged particle \cite{ref3}.

Between like particles, these correlated-fluctuation forces are 
attractive.  There is always a question, though, whether one is 
allowed to compute these attractive forces in molecular arrays as 
though they were the sum of individual rod-rod interactions.  In this 
paper we derive the interaction between pairs of cylindrical rods -- 
parallel or skewed -- in a salt solution.  We find that the conditions 
for the validity of pairwise additivity can be quite restrictive.  
Without the screening effect of the intervening salt the counterion 
correlation forces come from collective ionic fluctuations in the 
array that can not be decomposed into separate pairwise additive 
contributions.  The range of validity of different theories 
\cite{ref2,ref4} that formalize counterion correlation forces is 
practically limited.

We use the Lifshitz-Pitaevskii approach \cite{ref3,ref5} that begins 
with an artificial but easily formulated interaction between arrays of 
cylindrical rod-like molecules.  This approach implicitly reveals any 
dilute limit in which the pair potential between rods can be extracted 
from the array-array interaction.  Specifically, we examine the van 
der Waals attraction between two like anisotropic media, L and R, 
containing parallel cylindrical particles embedded in a medium m 
(Figure 1).  L and R are separated by an isotropic region of salt 
solution devoid of cylinders.

The anisotropic regions are composites of parallel cylindrical rods of 
radius $a$ at volume fraction $v = N~\pi a^{2}$; $N$ is the 
cross-sectional density of $N$ rods per unit area.  The cylinders c 
have anisotropic intrinsic susceptibilities $\epsilon_{\perp}^{c}$ and 
$\epsilon_{\parallel}^{c}$, perpendicular and parallel to the rod axis.  
For each region r =L, m, R the effective dielectric and ionic 
properties of the composites can be written in terms of their local 
values.  In this construction the axis parallel to the rod in region R 
is rotated about axis z to create an angle $\theta$ with respect to the rod 
x-axis in L.

Because ionic fluctuations are slow \cite{ref3}, we consider only 
'zero-frequency' van der Waals interactions, describing thermodynamic 
as opposed to quantum fluctuations.  The wave equations in such media 
follow
\begin{equation}
\nabla\left({\mbox{\boldmath $\epsilon$}}\nabla\phi({\bf r})\right) = 
k^{2}\phi({\bf r})
\end{equation}
where $k$ is the same as the Debye screening constant except for the 
missing dielectric constant $\epsilon$ in the denominator.  The 
dielectric susceptibility tensor ${\mbox{\boldmath $\epsilon$}}$ is 
taken only in the limit of zero frequency.  The susceptibility tensors 
${\mbox{\boldmath $\epsilon$}}_{r}$ for the composite regions r = L 
and R have diagonal elements 
$\epsilon_{x}^{r},~\epsilon_{y}^{r},~\epsilon_{z}^{r}$, and with the 
axes x and y of R rotated by an angle $\theta$.  For small enough 
volume fractions $v$, the components of the unrotated tensors for the 
composite media are \cite{ref6} $\epsilon^{L}$ and $\epsilon^{R}$
\begin{equation}
\epsilon_{\parallel} = \epsilon^{L}_{x} = \epsilon^{R}_{x} = 
\epsilon_{m}(1-v) + v\epsilon^{c}_{\parallel} = \epsilon_{m}(1 + 
v\Delta_{\parallel})
\end{equation}
\begin{equation}
\epsilon_{\perp} = \epsilon^{L}_{y} = \epsilon^{R}_{y} = 
\epsilon^{L}_{z} = \epsilon^{R}_{z} =  \epsilon_{m}\left(1 + 
\frac{2v\Delta_{\perp}€}{1-v\Delta_{\perp}€}\right) = 
\epsilon_{m}\left(\frac{1+v\Delta_{\perp}}{1-v\Delta_{\perp}}
\right)
\end{equation}
with
\begin{equation}
\Delta _{||}={{\varepsilon _{_{||}}^c-\varepsilon _m} \over 
{\varepsilon _m}}~~~~\  \Delta _\bot ={{\varepsilon _\bot ^c-
\varepsilon _m} \over {\varepsilon _\bot ^c+\varepsilon _m}}
\end{equation}
where $\epsilon_{m}$ is the isotropic susceptibility in region m and between 
cylinders in L and R; superscript c refers to the response of the 
cylinder material.  

The quantities $k_{r}^{2}$ in each region $r = L, m, R$ depend on 
volume averages of ionic strengths $n_{}=\ \sum\limits_Z^{} {Z^2}n_Z$ 
where $n_{Z}$ is the average number-density of ions of valence Z.  For 
region m, $k_m^2 = \frac{4\pi n_me^2}{kT}$, the Debye constant being defined in 
a standard way as $\kappa_{m}^{2}={{4\pi \kern 1pt n_me^2} \over 
{\varepsilon _m\kern 1pt kT}}$
	
For regions L and R, we distinguish the densities of mobile ions 
$n_{c}$ and $n_{m}$ in the cylinders and in the intervening space 
respectively.  The mobile ions associated with the cylinders model the 
tightly associated counterions where $n_{c}$ pertains to the average 
density of mobile charges within the cylinders.  The volume average 
$n_{L}$ and $n_{R}$ over the whole composite is $n_{L}€ = n_{R}€ = 
n_{c}€ v + n_{m}€ (1-v) = n_{m}€ + v (n_{c}€ - n_{m}€)$, which leads 
to
\begin{equation}
k_{L}^{2}€€=k_{R}^{2}= \frac{4\pi 
e^{2}€}{kT}\left(n_{m}+v(n_{c}-n_{m})\right) = k_{m}^{2}\left(
1 + v\frac{n_{c}-n_{m}}{n_{m}}\right).
\end{equation}
Because solutions of the wave equation require continuity of 
${\mbox{\boldmath $\epsilon$}}\nabla\phi$ perpendicular to the 
interfaces at $z = 0$ and $l$, it is the perpendicular response of 
$\epsilon^{L}$ and $\epsilon^{R}$ that couples with 
$k_{L}^{2}€€=k_{R}^{2}$ to create the ionic screening lengths in both 
semi-infinite regions.  For small enough volume fractions $v$
\begin{equation}
\frac{k_{L}^{2}€€}{\epsilon_{\perp}€} = \frac{k_{R}^{2}€€}
{\epsilon_{\perp}€} \approx \kappa_{m}^{2}\left( 1 + v\left(
\frac{n_{c}-n_{m}}{n_{m}}-2\Delta_{\perp}\right)\right).
\end{equation}€
From this point on we will simply write $\kappa^{2}$ (no subscript) for 
$\kappa_{m}€^{2}$.  

The 'zero-frequency' van der Waals interaction free energy per unit 
area between the two anisotropic semi-infinite regions L and R across 
a slab m of thickness $l$ is \cite{ref3}
\begin{equation}  
{\cal G}_{LmR}(l,\theta ) = {{kT} \over 2}\;\int
\frac{d^{2}{\bf Q}}{(2\pi )^{2}}~{\ln ({\cal D}(Q,l))\kern 1pt} = {{kT} \over {8\pi ^2}}\;\int_0^{2\pi }
 {d\psi \int_0^\infty  {\ln ({\cal D}(Q,l))\kern 1pt \kern 1pt Q\kern 1pt
  dQ}}
\end{equation}
Here, the $\bf Q$ are radial wave vectors within the $x,y$ plane; 
$\psi$ is for angular integration over all directions in $\bf Q$ to take 
account of anisotropy.  The secular determinant for the relevant modes 
can be written as
\begin{equation}
{\cal D}(Q,l) = 1-\Delta _{Lm}(\psi )\Delta _{Rm}(\theta -\psi )\times 
e^{-2\sqrt {Q^2+\kappa ^2}l}.
\label{blef}
\end{equation}
The functions $\Delta _{Lm}(\psi )$ and $\Delta _{Rm}(\theta -\psi ) $
necessarily go to zero as $N, v \longrightarrow 0$; both have the form 
(i = L, R)
\begin{equation}
\Delta _{im}(\varphi )\ =\ {{\varepsilon _m\sqrt {Q^2+\kappa ^2}-\ 
\varepsilon _\bot \sqrt {Q^2\left( {1+\left( {{\textstyle{{\varepsilon 
_{II}-\varepsilon _{_\bot }} \over {\varepsilon _{_\bot }}}}} 
\right)\cos ^2\varphi } \right)+{{k_i^2} \over {\varepsilon _{_\bot 
}}}}} \over {\varepsilon _m\sqrt {Q^2+\kappa ^2}+\ \varepsilon _\bot 
\sqrt {Q^2\left( {1+\left( {{\textstyle{{\varepsilon 
_{II}-\varepsilon _{_\bot }} \over {\varepsilon _{_\bot }}}}} 
\right)\cos ^2\varphi } \right)+{{k_i^2} \over {\varepsilon _{_\bot 
}}}}}}	
\end{equation}€
For small enough values of $v$, the $\Delta _{im}(\phi )=\Delta 
_{im}(\phi,v )$ functions can usually be expanded to terms linear in 
cylinder volume fraction $v$.  To lowest order in density, the 
interaction energy goes as $N^{2}$.  In this dilute limit, rods in the two 
media interact pairwise across the gap $l$.

Extraction of the pairwise interaction potential follows one of two 
procedures, depending whether or not the rods in L and R are parallel 
($\theta = 0$) or are skewed, to connect the per-unit-area interaction ${\cal 
G}(l,\theta)$ between planar regions of embedded cylinders with the 
pair interaction potential $g(l,\theta )$ between skewed cylinders, or 
the pair interaction energy per unit length $g(l,\theta = 0 )$ between 
parallel cylinders.  A sufficient condition for this connection is 
that ${\cal G}(l,\theta)$ be accurately expressed by the first, 
quadratic term in a series expansion in density $N$ or volume fraction 
$v$.  

The Pitaevskii {\sl ansatz} \cite{ref5} can be applied in two forms.  
For {\sl skewed} cylinders at a mutual angle $\theta$, the connection is
\begin{equation}
\left. \lim _{N\to 0}{{d^2{\cal G}(l,\theta )} \over 
{dl^2}}\right|_{l=R}€ = N^2\sin \theta \ 
g(R,\theta )\ +\ {\rm O}(N^3)	
\end{equation}
while for {\sl parallel} cylinders the connection is an Abel transform
\begin{equation}
\left.\lim _{N\to 0}{{d^2{\cal G}(l,\theta =0)} \over {dl^2}}\right|_{l=R}€ = 
N^2\int\limits_{-\infty }^{+\infty } {g(\sqrt {R^2+y^2},\theta =0)\ 
dy\ +\ {\rm O}(N^3)}.
\end{equation}
In both cases $R$ is the minimal separation between the cylinders.

Expanding ${\cal G}(l,\theta)$ to the second order in $v$ one obtains
\begin{equation}
{\cal G}(l,\theta) = -{{kT} \over 4\pi}\kappa ^2v^2\int_1^\infty {p\kern 
1pt dp\ e^{-2\kappa lp}}f(p,\theta )\ +{\rm O}(v^3)
\end{equation}
with
\begin{equation}
f(p,\theta ) = \frac{1}{2\pi}\int_{0}^{2\pi}\frac{\partial\Delta _{Lm}(\psi )}
{\partial v}\frac{\partial\Delta _{Rm}(\theta - \psi )}{\partial v} 
d\psi
\end{equation}
For {\sl skewed} cylinders this leads to
\begin{equation}
g(R,\theta ) = -\frac{kT}{\pi}{{\kappa ^4(\pi a^2)^2} \over {\sin \theta 
}}\int_1^\infty {p^3dp\ e^{-2\kappa Rp}}f(p,\theta )
\end{equation}
for {\sl parallel} cylinders through the inverse Abel transform one remains 
with
\begin{equation}
g(R,\theta =0)\,\ =\ \,-2{{kT} \over {\pi^{2}} }\kappa ^5(\pi 
a^2)^2\int_1^\infty {p^4dp\ e^{-2\kappa Rp}}K_0(2\kappa Rp)f(p,\theta 
=0)
\end{equation}
$K_{0}€(x)$ is the modified Bessel cylindrical function of order 0. 

Integrating over the azimuthal directions $\psi$ and taking the second 
derivative $\frac{d^{2}{\cal G}(l,\theta )}{dl^{2}}$, reveals three 
classes of interaction depending on the type of charge fluctuation: 
dipole-dipole, dipole-monopole, and monopole-monopole:
\begin{eqnarray}
\kern-40pt \left.{{{d^2 {\cal G}(l,\theta )} \over {dl^2}}} 
\right|_{D-D}\cong & & \ -\frac{kT}{\pi}\kappa ^4v^2\left[ {\Delta ^2_\bot +{1 \over 
4}\Delta _\bot \left( {\Delta _{II}\ -\ 2\Delta _\bot } \right)+\left( 
{\Delta _{II}\ -\ 2\Delta _\bot } \right)^2{{2\cos ^2\theta +1} \over 
{2^7}}} \right]\ \times \nonumber\\ & & \times \left\{ 
{{{6e^{-(2\kappa l)}} \over {(2\kappa l)^4}}\left( {1+(2\kappa 
l)+{{(2\kappa l)^2} \over 2}+{{(2\kappa l)^3} \over 6}} 
\right)-2{{e^{-(2\kappa l)}} \over {(2\kappa l)^2}}\left( {1+(2\kappa 
l)} \right)-Ei(-2\kappa l)} \right\}\nonumber
\end{eqnarray}
\begin{eqnarray}
\kern-40pt\left.{{{d^2{\cal G}(l,\theta )} \over {dl^2}}} \right|_{D-M}\cong & & \ 
2\frac{kT}{\pi}\kappa ^4v^2\left( {{{n_c-n_m} \over {2n_m}}} \right)\left[ {\Delta 
_\bot +{1 \over 8}\left( {\Delta _{II}\ -\ 2\Delta _\bot } \right)} 
\right]\ \left\{ {\left.  {{{e^{-(2\kappa l)}} \over {(2\kappa 
l)^2}}\left( {1+2\kappa l} \right)+Ei(-2\kappa l)} \right\}} 
\right.\nonumber
\end{eqnarray}
\begin{equation}
\left.  {{{d^2{\cal G}(l,\theta )} \over {dl^2}}} \right|_{M-M}\cong \ 
\frac{kT}{\pi}\kappa ^4v^2\left( {{{n_c-n_m} \over {2n_m}}} \right)^2Ei(-2\kappa 
l).
\end{equation}€

We examine the results for $g(R,\theta )$ and $g(R,\theta =0)$ in 
several limiting cases and then ask when one is permitted to use 
expressions that pre-suppose pairwise additivity of rod-rod 
interactions.

In the limit of {\sl vanishing salt concentration in the medium}, 
$2\kappa R \longrightarrow 0$,
\begin{eqnarray}
g(R,\theta )\ \cong & & \ -{3 \over {8\pi}}kT{{(\pi a^2)^2} \over {\sin 
\theta \ R^4}}\ \left[ {\Delta ^2_\bot +{1 \over 4}\Delta _\bot \left( 
{\Delta _{II}\ -\ 2\Delta _\bot } \right)+\left( {\Delta _{II}\ -\ 
2\Delta _\bot } \right)^2{{2\cos ^2\theta +1} \over {2^7}}} \right] 
\nonumber\\ & & \ +\ \frac{kT}{2\pi}{{(\pi a^2)^2} \over {\sin \theta \ R^2}}\left( 
{{{4\pi \kern 1pt n_ce^2} \over {\varepsilon _m\kern 1pt kT}}} 
\right)\left[ {\Delta _\bot +{1 \over 8}\left( {\Delta _{II}\ -2\Delta 
_\bot } \right)} \right]\ \nonumber\\ & & \ +\ \frac{kT}{\pi}{{(\pi a^2)^2} \over 
{\sin \theta }}\left( {{{4\pi \kern 1pt n_ce^2} \over {\varepsilon 
_m\kern 1pt kT}}} \right)^2\ln \left( {2\kappa R} \right)
\end{eqnarray}
\begin{eqnarray}
g(R,\theta =0)\ \cong & & \ -{9 \over {32\pi}}kT{{(\pi a^2)^2} 
\over {R^5}}\ \left[ {\Delta ^2_\bot +{1 \over 4}\Delta _\bot \left( 
{\Delta _{II}\ -\ 2\Delta _\bot } \right)+\left( {\Delta _{II}\ -\ 
2\Delta _\bot } \right)^2{3 \over {2^7}}} \right]\ + \nonumber\\ 
& & \ +\ {1 \over 
{4\pi}}kT{{(\pi a^2)^2} \over {R^3}}\left( {{{4\pi \kern 1pt n_ce^2} \over 
{\varepsilon _m\kern 1pt kT}}} \right)\left[ {\Delta _\bot +{1 \over 
8}\left( {\Delta _{II}\ -\ 2\Delta _\bot } \right)} \right]\ 
\nonumber\\ & & \ -\ 
\frac{kT}{2\pi}{{(\pi a^2)^2} \over R}\left( {{{4\pi \kern 1pt n_ce^2} \over 
{\varepsilon _m\kern 1pt kT}}} \right)^2.
\end{eqnarray}

When all salt concentrations equal zero, interactions come only from 
differences in dielectric susceptibility.  All monopolar terms vanish 
to leave only dipolar terms \cite{ref5}.

In the opposite case of {\sl strong ionic screening}, $2\kappa R \gg 1$, the 
only terms surviving are those of longest range that nevertheless 
decay exponentially.
\begin{eqnarray}
g(R,\theta )\ \cong & & \ -\frac{8}{\pi}kT{{\kappa (\pi a^2)^2} \over {\sin 
\theta }}{{e^{-2\kappa R}} \over {(2R)^3}}\ \left( {\Delta _{II}\ -\ 
2\Delta _\bot } \right)^2{{2\cos ^2\theta +1} \over {2^7}}\ 
\nonumber\\ & &  
-\frac{kT}{2\pi}{{\kappa ^2(\pi a^2)^2} \over {\sin \theta }}{{e^{-2\kappa R}} 
\over {(2R)^2}}\left( {\Delta _{II}\ -\ 2\Delta _\bot } \right)\left( 
{{{n_c-n_m} \over {2n_m}}} \right)\ \nonumber\\ & &\ -\frac{kT}{\pi}{{\kappa ^3(\pi a^2)^2} 
\over {\sin \theta }}{{e^{-2\kappa R}} \over {(2R)}}\ \left( 
{{{n_c-n_m} \over {2n_m}}} \right)^2
\end{eqnarray}
\begin{eqnarray}
g(R,\theta )\ \cong & &  -16{{kT} \over {\pi^{2}} }\kappa ^5(\pi a^2)^2
\sqrt{{2 \over 
\pi }}\ {{e^{-2\kappa R}} \over {(2\kappa R)^{{7 \over 2}}}}\left[ 
{\Delta ^2_\bot +{1 \over 4}\Delta _\bot \left( {\Delta _{II}\ -\ 
2\Delta _\bot } \right)+\left( {\Delta _{II}\ -\ 2\Delta _\bot } 
\right)^2{3 \over {2^7}}} \right]\ \nonumber\\ & & \ \ \ \ \ + 8{{kT} 
\over {\pi^{2}} 
}\kappa ^5(\pi a^2)^2\sqrt {{2 \over \pi }}\ {{e^{-2\kappa R}} \over 
{(2\kappa R)^{{5 \over 2}}}}\left( {\Delta _\bot \ +\ {1 \over 
{16}}(\Delta _{II}\ -\ 2\Delta _\bot )} \right)\left( {{{n_c-n_m} 
\over {2n_m}}} \right)\ \nonumber\\ & &\ \ \ -2{{kT} \over {\pi^{2}} }\kappa ^5(\pi 
a^2)^2\sqrt {{2 \over \pi }}\ {{e^{-2\kappa R}} \over {(2\kappa R)^{{3 
\over 2}}}}\ \left( {{{n_c-n_m} \over {2n_m}}} \right)^2
\end{eqnarray}
When all $\epsilon $'s are {\sl equal} but there are differences in ionic 
densities $n_{c}€$ and $n_{m}€$, only monopolar terms endure,
\begin{equation}
g(R,\theta )\ \cong \ -\frac{kT}{\pi}{{\kappa ^4(\pi a^2)^2} \over {\sin \theta 
}}{{e^{-2\kappa R}} \over {(2\kappa R)}}\left( {{{n_c-n_m} \over 
{2n_m}}} \right)^2
\end{equation}
\begin{equation}
g(R,\theta =0)\ \cong \ -2{{kT} \over {\pi^{2}} }\kappa ^5(\pi a^2)^2\sqrt 
{{2 \over \pi }}{{e^{-2\kappa R}} \over {(2\kappa R)^{{3 \over 
2}}}}\left( {{{n_c-n_m} \over {2n_m}}} \right)^2.
\end{equation}€

There is a curious further limit -- {\sl unscreened monopole-monopole 
correlated fluctuations} -- where $n_{m} \longrightarrow 0$ but 
cylinders still carry counterions $n_{c}$,
\begin{equation}
g(R,\theta )\cong \frac{kT}{\pi}{{(\pi a^2)^2} \over {\sin \theta }}\left( 
{{{4\pi \kern 1pt n_ce^2} \over {\varepsilon _m\kern 1pt kT}}} 
\right)^2\ln \left( R \right)\ 
	\label{crap}
\end{equation}
(to within additive constants) and
\begin{equation}
g(R,\theta =0)\cong -\frac{kT}{2\pi}{{(\pi a^2)^2} \over R}\left( {{{4\pi \kern 1pt 
n_ce^2} \over {\varepsilon _m\kern 1pt \kern 1pt kT}}} \right)^2.
	\label{oosawa}
\end{equation}€

The $n_{m}€= 0$ form for parallel rods has been obtained analytically 
\cite{ref4} before and checked against computer simulation 
\cite{ref2}.  Our result Eq.\ref{oosawa} however indicates that only the 
linearized form of the Oosawa potential is pairwise additive in an 
array, i.e.  there is no leveling off of the correlation attraction as 
the concentration of mobile counterions, $n_{c}€$ in our model, increases.

That Eqs.  \ref{crap} and \ref{oosawa} indeed correspond to monopolar 
fluctuations can be established by recalling \cite{ref7} monopolar 
charge fluctuation forces between small spheres of radius $a$ at 
separation $R$ that go as $\sim kT \left(\frac{a}{R}\right)^{2} 
~e^{-2\kappa R}$.  Assuming that this ion correlation potential sums 
pairwise between different parts of the rods in either skewed or 
parallel configuration, one obtains the forms of Eqs.  \ref{crap} and 
\ref{oosawa}.

When can one use pairwise rod-rod interactions in solutions and in 
arrays?  

The central requirement of the Pitaevskii construction is to 
create composite media whose dielectric properties are linear in the 
number density of interacting particles.  The interaction between 
composites then goes as the square of the number densities.  With salt 
solutions, requisite dependence on the square of densities requires 
more than linear dielectric response.

Expanding the dispersion relation ${\cal D}(Q,l)$ (Eq. \ref{blef}) in 
powers of cross sectional density of the cylindrical array, we see 
that the expansion
\begin{equation}
\Delta _{Lm}(\psi )\Delta _{Rm}(\theta -\psi )\ \cong \ v^2\ \left.  
{{{\partial \Delta _{Lm}(\psi )} \over {\partial v}}} 
\right|_{v=0}\left.  {{{\partial \Delta _{Rm}(\theta -\psi )} \over 
{\partial v}}} \right|_{v=0}\ +\ O(v^3)
\end{equation}
only makes sense if two inequalities, 
\begin{equation}
v\kern 1pt \kern 1pt {{(n_c-\  n_m)} \over {n_m}}\  <<\  1 ~~~~
v	\Delta_{\perp} \  <<\  1
	\label{trap}
\end{equation}€
are strictly satisfied. 

Whenever salt concentration $n_{m} \longrightarrow 0$ while the 
concentration of the mobile charges within the cylinders $n_{c}€$ remains 
finite, expansion in $N $ must always violate the first inequality.  
Bluntly, limiting forms such as Eqs.  \ref{crap} and \ref{oosawa} 
cannot be used in problems where there are more than two molecules.  
Their very range carries within it the source of their 
self-invalidation as pairwise-additive interactions.

Traditional van der Waals forces, due to differences in dielectric 
susceptibility, will often not be pairwise additive.  For 
low-frequency fluctuations, the second inequality in Eq.  \ref{trap} 
can easily be violated \cite{ref5}.  Finite values of the medium salt 
concentration always create an exponentially damped contribution when 
$2\kappa_{m} R \gg 1$ .  Though screening severely weakens forces, it 
does act to make pairwise summation more accurate.

Charge fluctuation forces in condensed systems can be qualitatively 
non-additive.  This feature imposes questions that are sometimes 
neglected in analyses of colloidal and macromolecular systems.  When 
addressing correlated ionic fluctuations it appears to be necessary to 
think in terms of collective excitations that extend over all 
molecules in an array.
%--------------------------------------------------------------------------

\vfill
\eject

\begin{figure}
\begin{center}
%\epsfxsize=15cm
%\epsfysize=13cm
%\epsffile{figure.eps}
\caption{Array-interaction geometry.  Region L is confined 
to $z < 0$, region R to $z > l$; isotropic region m is the slab in 
between.  The cylindrical array in R is rotated about the z axis by an 
angle $\theta$ with respect to L. Not only the dielectric constants of the 
composite media R and L but also the dielectric constants of the 
cylindrical rods themselves are anisotropic. 
$\epsilon_{\parallel}^{c},~\epsilon_{\perp}^{c},~n_{c}$ refer to the 
cylinder material.}
\end{center}
\end{figure}

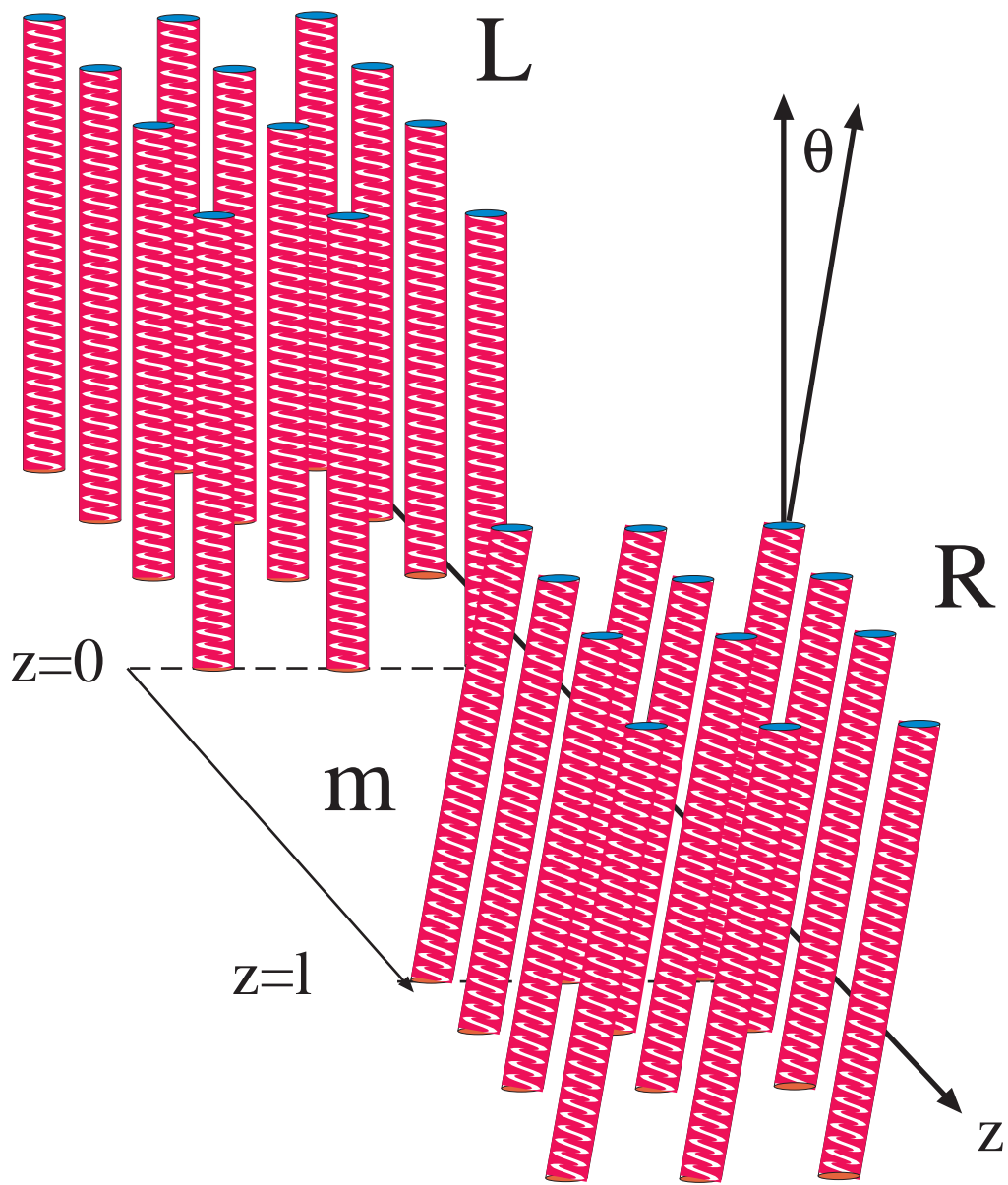
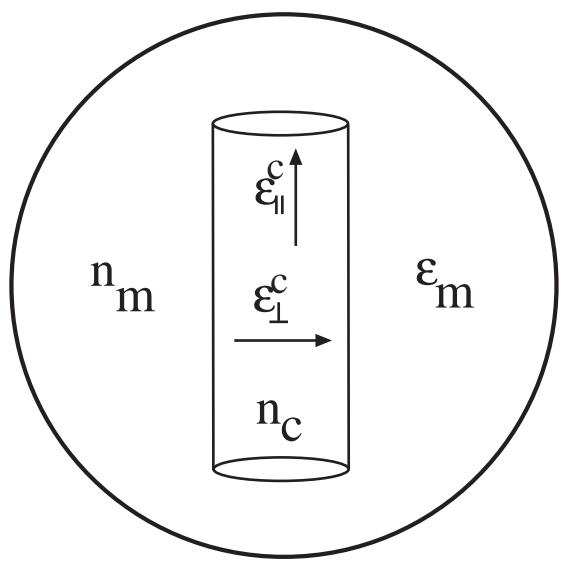
\end{document}